\documentstyle[11pt,newpasp,twoside,epsf]{article}
\def\edcomment#1{\iffalse\marginpar{\raggedright\sl#1\/}\else\relax\fi}
\reversemarginpar

\def\arcsec{\ifmmode '' \else $''$\fi}
\def\arcmin{\ifmmode ' \else $'$\fi}
\def\arcsecpoint{\ifmmode ''\!. \else $''\!.$\fi}
\def\arcminpoint{\ifmmode '\!. \else $'\!.$\fi}
\def\kms{\ifmmode {\rm km\ s}^{-1} \else km s$^{-1}$\fi}
\def\Hubble{\ifmmode {\rm km\ s}^{-1}\ {\rm Mpc}^{-1} 
	\else km s$^{-1}$ Mpc$^{-1}$\fi}
\def\ergsec{\ifmmode {\rm ergs\ s}^{-1} \else ergs s$^{-1}$\fi}
\def\eflux{\ifmmode {\rm ergs\ s}^{-1}\;{\rm cm}^{-2}
	  \else ergs s$^{-1}$ cm$^{-2}$\fi}
\def\efluxA{\ifmmode {\rm ergs\ s}^{-1}\;{\rm cm}^{-2}\;{\rm \AA}^{-1}
	  \else ergs s$^{-1}$ cm$^{-2}$ \AA$^{-1}$\fi}
\def\efluxHz{\ifmmode {\rm ergs\ s}^{-1}\;{\rm cm}^{-2}\;{\rm Hz}^{-1}
	  \else ergs s$^{-1}$ cm$^{-2}$ Hz$^{-1}$\fi}
\def\cc{\ifmmode {\rm cm}^{-3} \else cm$^{-3}$\fi}
\def\vFWHM{\ifmmode v_{\mbox{\tiny FWHM}} \else
            $v_{\mbox{\tiny FWHM}}$\fi}
\def\Msun{\ifmmode M_{\odot} \else $M_{\odot}$\fi}
\def\Lsun{\ifmmode L_{\odot} \else $L_{\odot}$\fi}
\def\Halpha{\ifmmode {\rm H}\alpha \else H$\alpha$\fi}
\def\Hbeta{\ifmmode {\rm H}\beta \else H$\beta$\fi}
\def\Hgamma{\ifmmode {\rm H}\gamma \else H$\gamma$\fi}
\def\Hdelta{\ifmmode {\rm H}\delta \else H$\delta$\fi}
\def\Lya{\ifmmode {\rm Ly}\alpha \else Ly$\alpha$\fi}
\def\Lyb{\ifmmode {\rm Ly}\beta \else Ly$\beta$\fi}
\def\hi{\ifmmode {\rm H}\,{\sc i} \else H\,{\sc i}\fi}
\def\hii{\ifmmode {\rm H}\,{\sc ii} \else H\,{\sc ii}\fi}

\def\ciii{\ifmmode {\rm C}\,{\sc iii} \else C\,{\sc iii}\fi}
\def\civ{\ifmmode {\rm C}\,{\sc iv} \else C\,{\sc iv}\fi}

\begin{document}
\title{Spectropolarimetry of Broad \Halpha ~Lines and Geometry of the BLR}
\author{Marshall H. Cohen}
\affil{California Institute of Technology, Pasadena, CA 91125}
\author{Andr\'{e} R. Martel}
\affil{Johns Hopkins University, 3400 N. Charles Street, Baltimore, MD 21218}

\begin{abstract}
   In a small fraction of Broad Line Radio Galaxies (BLRG) and Seyfert
1 galaxies, the polarization position angle rotates across the broad
emission lines, especially at \Halpha.  An understanding of this
behavior can potentially yield important information on the scattering
geometry in the nucleus.  We show two examples of this phenomenon, 3C
445, a BLRG, and Mrk 231, a Seyfert 1, and present an equatorial
scattering model that explains some of its features in a
straightforward way.  \end{abstract}

\section{Introduction}

  The literature contains a number of examples of rotation of the
polarization position angle ({\em PA}) in \Halpha ~in Seyfert 1 galaxies and
in broad-line radio galaxies (BLRG). This has usually been explained
as radiation from the nuclear sources (continuum and BLR) being
scattered from separated clouds, so that different Doppler-shifted
wavelengths are seen at different {\em PA}.  The most elaborate discussion
of this is by Martel (1998) who decomposed \Halpha ~in NGC 4151 into
components each with its own {\em PA}. Several of the components were able
to be associated with various features in the galaxy.  Such an
interpretation, dominated by radial motions, suffers from a lack of
knowledge of the relative location and velocities of the emitters and
scatterers.

  In this paper we investigate a different possibility, that the
rotations are caused by scattering of \Halpha ~radiation on nearby
clouds which see both red- and blue- shifted radiation but from
different directions.  The scattering planes are different for the red-
and blue- shifted rays and an integration gives a {\em PA} that can rotate
through \Halpha. A discussion of this idea is in Cohen et al. (1999;
hereafter C99) and it is mentioned by Goodrich and Miller (1994).

\section{Examples}

\begin{figure}[t]

\plotfiddle{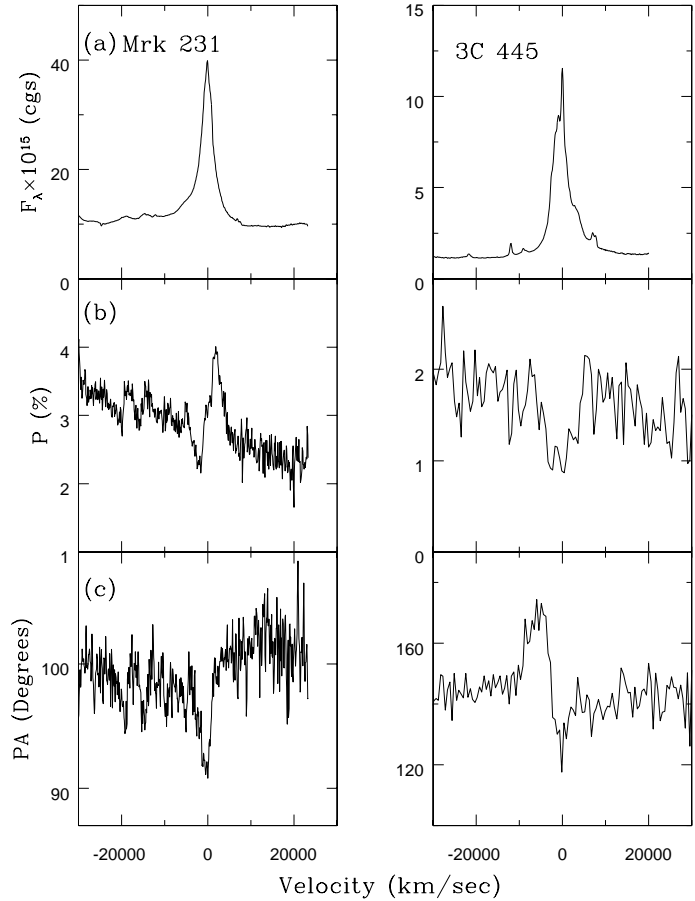}{4.5in}{0}{48}{48}{-144}{-20}

\caption{\Halpha ~spectral region of Mrk 231 and 3C 445~: (a) total
flux (b) percentage polarization (c) {\em PA}.  Note the S-shaped
changes in {\em P} in Mrk 231 and in {\em PA} in 3C 445.}
\end{figure}

\subsection{Mrk 231}

   Figure 1 shows (a) total flux, (b) polarization {\em P}, and (c)
position angle {\em PA} for the \Halpha ~region of Mrk 231 (Martel
1996; hereafter M96) and 3C 445 (C99).  Note that {\em P} is S-shaped
in Mrk 231, decreasing on the blue side and increasing on the red side
of \Halpha.  This shows that the {\em PA} of the \Halpha ~flux is
different in the two wings; and that either {\em P} is higher in
\Halpha ~than in the continuum, or (more likely) that the dilution of
{\em P} by unpolarized components is lower in (continuum + \Halpha)
than in the continuum alone.  In (c), it can be seen that the {\em PA}
of \Halpha ~is similar to that of the continuum in most of the red
wing, but {\em PA}(blue) is smaller. In fact the intrinsic
(continuum-corrected) shift in {\em PA}(blue) must be rather
substantial, because in (b) the dip in the blue is nearly as big as
the rise in the red.  M96 modelled this line with 5 Gaussian
components, each with its own wavelength and {\em PA}.  His result is
that there is a trend in {\em PA} corresponding to velocity, and the
{\em PA} difference between the red and blue wings is about 30\deg.

\subsection{3C 445}

   Spectra for the BLRG 3C 445 are shown in the right-hand panels of
Figure 1 (C99).  The polarization fraction {\em P} has a weak S pattern, but
{\em PA} has striking changes across \Halpha. A decomposition of the flux
into line and continuum components (C99) shows that {\em PA}(\Halpha) has
about a 45\deg ~change from the red to the blue wing, and much of the
change occurs about 2500 \kms ~to the blue of the line center.

\subsection{Others}

   The \Halpha ~polarization in seven Sey 1 galaxies is discussed in
M96.  For most, the continuum and line polarizations are low,
$\sim1$\%, and they all show structure across their \Halpha ~profiles
in both {\em P} and {\em PA}.  Mrk 304 and NGC 3516 are similar to Mrk
231 in having a strong S shape in {\em P}, and the S pattern is offset
to the blue by perhaps 2500 \kms ~in Mrk 304. As for Mrk 231, the {\em
PA} also changes where {\em P} dips.  In Mrk 704, the {\em PA} rotates
about 8\deg ~across the central 5500 \kms ~of \Halpha.  A few BLRG, in
addition to 3C 445, show polarization changes in \Halpha.  3C 227 has
both \Halpha ~and \Hbeta ~offset by $\sim20$\deg ~from the continuum;
in this case the {\em PA} of the line radiation itself is
approximately constant (C99).

   Many broad-absorption-line quasars show {\em PA} rotations (Ogle
1998, Ogle et al. 1999), especially in the deep absorption troughs.
It is not clear if these rotations are generically similar to those
seen in the emission lines of BLRG and Seyfert 1's.

\section{An Equatorial Model} 

\begin{figure}[t]

\plotfiddle{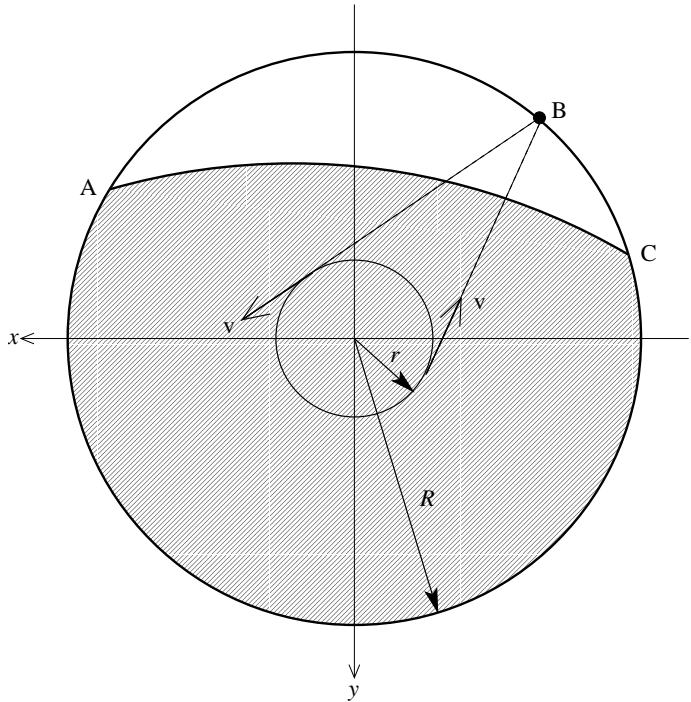}{4.0in}{0}{50}{50}{-144}{20}

\caption{Equatorial plane of an AGN viewed along the $z$-axis. \Halpha
~clouds orbit with velocity $v$ on the inner circle.  Emission from the
clouds is scattered by material on the outer circle and is seen by an
observer at angle $\theta$ from the $z$-axis. A dusty torus hides the
hatched region from direct view. At point B, the scattering planes are
different for the blue and red rays, and the observed blue and red
polarizations are at different {\em PA}'s.}

\end{figure}

   Figure 2 shows the equatorial plane of an idealized AGN.
\Halpha-emitting clouds orbit with velocity $v$ on the inner circle of
radius $r$, and scattering clouds lie on the outer circle of radius $R$
($\rho = r/R$).  Both the \Halpha ~and scattering clouds are uniformly
distributed around the circles, and the former radiate isotropically.
The observer is in the $y-z$ plane at angle $\theta$ from the
$z$-axis.  AGN unification scenarios generally assume that BLRG are
seen near the boundary of an obscuring dusty torus, which shields the
central quasar from direct view. In Figure 2, the projection of the
torus' edge on the equatorial plane is represented by arc AC and the
dashed region is hidden, and the observer has an asymmetric view of the
equatorial plane.  The asymmetry is necessary to obtain rotation of the
{\em PA}, and also for the model to mimic the observed spread between
the optical continuum {\em PA} and the radio axis, which we assume to
be projected onto the $y$ axis.  Point B receives both red- and
blue-shifted radiation but the scattering planes are different so the
red and blue wings of the scattered line will be at different position
angles. When this is integrated over the visible arc ABC, the {\em PA}
can rotate as a function of velocity.

   Dust is much more efficient than electrons at scattering and so we
take $R$ to be at the inner wall of the dusty torus. If this is near 1
pc and the BLR is at a few light months, then $\rho \sim 0.1$.  This
picture is very simplified since in reality the torus will be irregular
(think of the Galactic Center).  It would be more realistic to put
clouds randomly near the outer circle and to vary $R$; and also of
course to add structure in the $z$ direction for both the emission and
scattering clouds. These elaborations are not warranted at present.

   In the examples shown below, we have taken $\rho = 0.1$ and the only
free parameters are $\theta$ and the points A and C which define the
visible arc.  In fact, a wide range of results can be obtained by
varying A and C, essentially because the integration adds together many
vectors with a wide range of angles, and the (often small) residual
vector can depend critically on the details of cancellation.  For the
results shown here we weigh radiation in the equatorial plane by the
inverse square of the distance and with a cosine that makes the
opposite point invisible.

   Light from the central continuum source, which has no motions in our
model, will also be scattered from the arc ABC, and its {\em PA} will
be the same as that of the zero-velocity point of \Halpha. However, in
some AGN these {\em PA}'s differ. This can be accomodated in our model
by adding another velocity, an outflow of \Halpha, or circular or
radial motions of the scatterers. This opens up a wide range of
parameter space which we have explored only in a limited fashion. Here,
we only discuss models where the \Halpha ~clouds have circular motions
and the scatterers are stationary.

   If the arc ABC is strictly symmetric around the $x$ or $y$ axis then
the continuum polarization will be parallel or perpendicular to the
radio axis, respectively. But we expect that in general the $x$ axis
will be obscured and the arc will be more or less as shown. This means
that radio-optical {\em PA} differences will avoid 0\deg, but values
near 90\deg ~will not be common either. This is in accord with
observations of BLRG; for five BLRG in C99 the median difference is
31\deg. If we take the radio axis to be given by the smallest-scale
observations with the VLBA, and the optical {\em PA} to be that of the
continuum near \Halpha, then the radio-optical differences for Mrk 231
and 3C 445 are 33\deg ~and 31\deg, respectively (M96; Ulvestad et al.
1999; C99)

\section{Results} 

\begin{figure}[t]

\plotfiddle{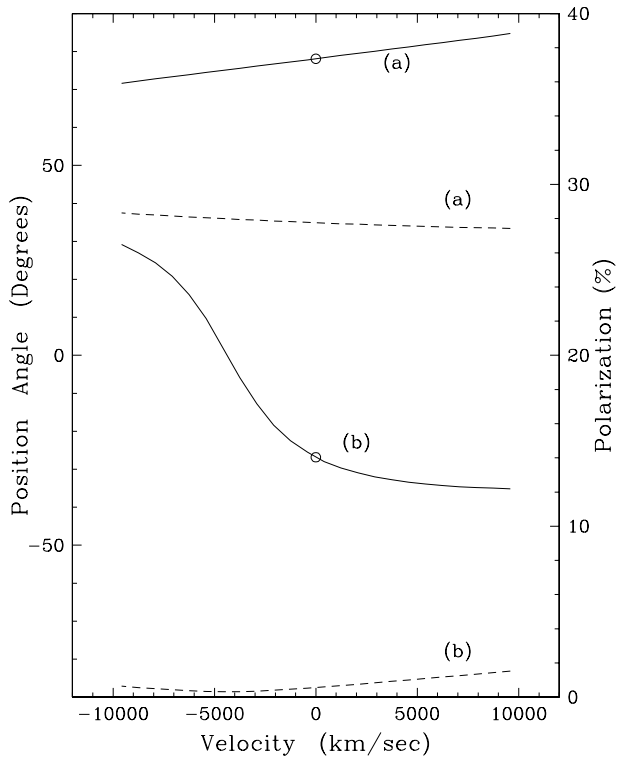}{4.0in}{0}{40}{40}{-144}{-20}

\caption{Calculations based on the model sketched in Figure 2 with
$\rho = 0.1$, $\theta = 20\deg$, and $v=10,000$ \kms. Solid curves are
the position angle, and dotted curves are the percentage polarization,
of the scattered \Halpha ~radiation. (a) arc ABC is from 1.1$\pi$ to
$1.8\pi$; (b) arc ABC is from 1.05$\pi$ to 2.0$\pi$.  The open circles
show the {\em PA} of the continuum.}

\end{figure}

   Figure 3 shows two examples of the many possibilities that can be
obtained with the model. To compare them with observations (Fig. 1) it
is necessary to combine \Halpha ~and continuum components.  In case
(a), the {\em PA} rotates uniformly and the continuum {\em PA} is in
the middle of the \Halpha ~range. This results in an S-shaped {\em PA}
curve as in 3C 445. However, two features of the observation are
inconsistent with the model: the measured curve is symmetric around
-2500 \kms ~instead of 0 \kms, and the radio-optical {\em PA}
difference is 31\deg, rather far from the model 75\deg. These
discrepancies, especially the first, appear to be rather difficult to
resolve; but adding a velocity, e.g. circular motion of the
scatterers, introduces many possibilities including offsetting the
{\em PA} of the zero-velocity component from the continuum.  The
polarization in the scattered light is high, nearly 30\%, which is
good because there also could be non-scattered components of both
\Halpha ~and continuum, which would dilute the polarization.

   In case (b) the continuum {\em PA} is close to that of the red wing
of \Halpha, and this gives {\em P} and {\em PA} curves like those in
Mrk 231.  The model radio-optical {\em PA} difference is 27\deg, close
to the observed value of 33\deg.  However, the cancellation of the
vectors is nearly complete, and the net polarization is only about 1\%.
The observed polarization is 3\%, and even without dilution the model
produces insufficient polarization.  This is a serious consideration,
since we do expect that there will be other components to the continuum
and line radiation.  There might be attenuated and reddened light seen
directly fron the BLR and the central continuum source; C99 estimate
that for 3C 445 the central source is seen through about 2 magnitudes
of extinction.


   We conclude that the equatorial scattering model in Figure 2 can
explain some of the features seen in Type 1 AGN polarization data.
Discrepancies remain but these may be resolved in the future by
exploring the parameter space of radial and circular motions of the
emitters and scatterers.

\end{document}